\documentclass[a4paper]{article}

\usepackage{INTERSPEECH2019}
\usepackage[x11names]{xcolor}
\usepackage{acronym}
\usepackage{framed}
\usepackage{array}
\usepackage{xspace}
\usepackage{enumitem}

\newacro{tts}[TTS]{text-to-speech}
\newcommand{\tts}{\ac{tts}\xspace}
\newacro{mos}[MOS]{mean opinion score}
\newcommand{\mos}{\ac{mos}\xspace}
\newacro{sxs}[SxS]{side by side}
\newcommand{\sxs}{\ac{sxs}\xspace}

\newcommand{\rp}{\textbf{R\textsuperscript{p}}}
\newcommand{\ri}{\textbf{R\textsuperscript{i}}}
\newcommand{\rOne}{\textbf{R\textsuperscript{1}}}
\newcommand{\tp}{\textbf{T\textsuperscript{p}}}
\newcommand{\ti}{\textbf{T\textsuperscript{i}}}
\newcommand{\tOne}{\textbf{T\textsuperscript{1}}}
\newcommand{\tTwo}{\textbf{T\textsuperscript{2}}}
\newcommand{\textOne}{\textbf{Text\textsuperscript{1}}}
\newcommand{\rfOne}{\textbf{R\textsubscript{F1}}}
\newcommand{\rmOne}{\textbf{R\textsubscript{M1}}}
\newcommand{\rfTwo}{\textbf{R\textsubscript{F2}}}
\newcommand{\rmTwo}{\textbf{R\textsubscript{M2}}}
\newcommand{\tfOne}{\textbf{T\textsubscript{F1}}}
\newcommand{\tfTwo}{\textbf{T\textsubscript{F2}}}
\newcommand{\tmOne}{\textbf{T\textsubscript{M1}}}
\newcommand{\tmTwo}{\textbf{T\textsubscript{M2}}}

\definecolor{my_green}{RGB}{184, 224, 164}
\definecolor{my_yellow}{RGB}{255, 230, 179}

\title{Evaluating Long-form Text-to-Speech: Comparing the Ratings of Sentences and Paragraphs}
\name{Rob Clark, Hanna Silen, Tom Kenter, Ralph Leith}
\address{Google U.K.}
\email{rajclark@google.com, silen@google.com, tomkenter@google.com, leith@google.com}

\begin{document}

\maketitle
\begin{abstract}
Text-to-speech systems are typically evaluated on single sentences.
When long-form content, such as data consisting of full paragraphs or dialogues is considered, evaluating sentences in isolation is not always appropriate as the context in which the sentences are synthesized is missing.

In this paper, we investigate three different ways of evaluating the naturalness of long-form text-to-speech synthesis.
We compare the results obtained from evaluating sentences in isolation, evaluating whole paragraphs of speech, and presenting a selection of speech or text as context and evaluating the subsequent speech.
We find that, even though these three evaluations are based upon the same material, the outcomes differ per setting, and moreover that these outcomes do not necessarily correlate with each other.
We show that our findings are consistent between a single speaker setting of read paragraphs and a two-speaker dialogue scenario.
We conclude that to evaluate the quality of long-form speech, the traditional way of evaluating sentences in isolation does not suffice, and that multiple evaluations are required.
\end{abstract}

\section{Introduction}
\label{sec:introduction}

Traditionally, \tts systems are trained on corpora of isolated sentences.
As such, their output is optimized, if only indirectly and inadvertently, for synthesizing isolated sentences.
As the use of \tts proliferates and the application of \tts extends into domains where the required output is high quality discourse, long-form (multi-sentence) data is being used more frequently to build voices and to evaluate the quality of the long-form output.

The traditional evaluation approaches used in \tts are designed to assess the quality of synthesized sentences in isolation using metrics such as \mos \cite{p800.1} and \sxs \footnote{ Also referred to as AB tasks.} discriminative tasks.
For long-form \tts, i.e., speech passages longer than one sentence, this evaluation scenario is limited in terms of what it can be used to evaluate; presenting sentences in isolation means that they are being evaluated out of their natural context. Long-form speech---which may consist of either single speaker data, such as an audio book, a news article, or a public speech; or multi-speaker data such as a conversation between multiple participants---should ideally be evaluated as a whole, because evaluating the quality of isolated sentences will not inform us of the overall quality of the discourse experience, which includes factors such as the appropriateness of prosody in context and fluency at paragraph-level.

The most obvious approach to evaluate long-form \tts is to use the existing standard evaluation techniques and simply present whole paragraphs or dialogues to raters.
Doing so, however, raises questions about the impact of providing longer stimuli that vary in length, both from the perspective of increasing the cognitive load of the raters through presenting them with more material, as from the perspective of increasing the overall variability in the length of stimuli. Including paragraph length as a factor in any subsequent analysis is often impractical as it drastically increases the amount of evaluation material required to fully control for it and still obtain a meaningful result. 

An additional scenario, which sits between evaluating isolated sentences and full long-form passages, would be to evaluate the quality of passages of speech in their immediate context.
In this scenario, full long-form passages are divided into two parts to form a context part and a stimulus part.
Raters are asked to evaluate the quality of the speech stimulus part as a continuation of a given context part, and are presented with the speech (or, potentially, just the text) of the context immediately before hearing the stimulus. Our  hypothesis is that we can achieve a higher sentence-level precision in this scenario than we could if the sentences were presented in isolation---as listeners are explicitly asked to evaluate whether the stimulus is appropriate for a specific context rather than being allowed to hypothesize a context for which the stimulus would be appropriate---while keeping the cognitive load for raters low compared to presenting them with full paragraphs.

To develop a better understanding of the potentials of the methods described above:
\begin{itemize}[leftmargin=3ex]
    \item We analyze three different ways of evaluating long-form \tts speech. To the best of our knowledge this is the first time a formal comparison based on a multitude of experiments has been performed;
    \item We show that both evaluating long-form \tts speech as paragraphs and as context-stimulus pairs yields results distinctly different from the traditional single sentence evaluation approach, which is remarkable given that the evaluations in all settings are based on the same material;
    \item We propose to combine these evaluations to get the most complete picture of long-form \tts quality.
\end{itemize}

As we are interested primarily in the relative differences of results between the various evaluation scenarios, rather than the relative differences between the \tts systems used, we focus on \mos tasks in this paper, and leave out \sxs evaluations. 

The remainder of this paper is organized as follows: Section~\ref{sec:related} discusses related work and existing approaches to (long-form) \tts evaluation. Section~\ref{sec:ways} details the three ways of evaluating long-form \tts that we propose.
Experimental details are presented in Section~\ref{sec:experimental_setup}. Sections~\ref{sec:results}~and~\ref{sec:further_analyis} present the results of the main and additional experiments, respectively. Section~\ref{sec:conclusions} concludes.

\section{Related Work}
\label{sec:related}

\colorlet{examplecolor}{my_green}
\colorlet{contextcolor}{my_yellow}

\colorlet{framecolor}{white}
\colorlet{unusedframecolor}{gray!30!white}

\setlength{\FrameSep}{2pt}

\newenvironment{example_stimulus}{%
 \def\FrameCommand{\fboxrule=\FrameRule\fboxsep=\FrameSep \fcolorbox{framecolor}{examplecolor}}%
 \MakeFramed {\FrameRestore}}%
{\endMakeFramed}

\newenvironment{context_stimulus}{%
 \def\FrameCommand{\fboxrule=\FrameRule\fboxsep=\FrameSep \fcolorbox{framecolor}{contextcolor}}%
 \MakeFramed {\FrameRestore}}%
{\endMakeFramed}

\newenvironment{unused_stimulus}{%
\def\FrameCommand{\color{gray}\fboxrule=\FrameRule\fboxsep=\FrameSep \fcolorbox{unusedframecolor}{white}}%
\MakeFramed {\FrameRestore}}%
{\endMakeFramed}

\def\textone{\footnotesize When former paratrooper and helicopter mechanic Adam Ely offered to fix his daughter's friend's car, he had what he calls ``a light bulb moment''.}
\def\texttwo{\footnotesize "It was super easy to do, I saved her at least \$80, and I thought, `I'd like to do more of this'," Adam, from Oklahoma, told the BBC.}
\def\textthree{\footnotesize Feeling inspired to help more people in need, Adam and his wife, Toni, set up Hard Luck Automotive Services (HLAS) in 2017.}

\def\examplettone{%
\begin{example_stimulus}%
\textone%
\end{example_stimulus}%
}

\def\contexttone{%
\begin{context_stimulus}%
\textone%
\end{context_stimulus}%
}

\def\unusedtone{%
\begin{unused_stimulus}%
\textone%
\end{unused_stimulus}%
}

\def\examplettwo{%
\begin{example_stimulus}%
\texttwo%
\end{example_stimulus}%
}

\def\contextttwo{%
\begin{context_stimulus}%
\texttwo%
\end{context_stimulus}%
}

\def\exampletthree{%
\begin{example_stimulus}%
\textthree%
\end{example_stimulus}%
}

\def\unusedtthree{%
\begin{unused_stimulus}%
\textthree%
\end{unused_stimulus}%
}

\def\contexttonetwo{%
\begin{context_stimulus}%
\textone
\newline\newline\vskip-0.75ex
\texttwo
\end{context_stimulus}%
}

\def\examplep{%
\begin{example_stimulus}%
\textone%
\newline\newline\vskip-1ex
\texttwo%
\newline\newline\vskip-1ex
\textthree%
\end{example_stimulus}%
}

\begin{figure*}[ht]
    \centering
\begin{minipage}[t][4.7cm][c]{5cm}
    \begin{center}(a)\end{center}
    \contexttone
    \examplettwo
    \unusedtthree
\end{minipage}
\hspace{1ex}
\begin{minipage}[t][4.7cm][c]{5cm}
    \begin{center}(b)\end{center}
    \unusedtone
    \contextttwo
    \exampletthree
\end{minipage}
\hspace{1ex}
\begin{minipage}[t][4.7cm][c]{5cm}
    \begin{center}(c)\end{center}
    \contexttonetwo
    \exampletthree
\end{minipage}
    \caption{Illustration of three ways to evaluate single sentences that are part of a three sentence paragraph, using other parts of the paragraph as context. Green boxes contain the audio to be evaluated. Yellow boxes are sentences presented as context (text and/or audio), not to be evaluated. White boxes show sentences of the paragraph not used in the rating task. (a) and (b) present the single previous sentence as context, while (c) presents two previous sentences in the paragraph.
    {\scriptsize(Text courtesy of BBC News)}}
    \label{fig:eval_isolated_with_context}
\end{figure*}

The currently used \mos \cite{p800.1} and \sxs tasks for evaluating TTS naturalness were established in \cite{van1995quality,campbell2007evaluation}.
Extensions and improvements to \mos evaluation have been made previously \cite{viswanathan2005measuring, wester2015we, shirali2018mos}, but none of this work covers the long-form scenario.

In \cite{mendelson2017beyond}, the point is made that evaluating sentences in isolation when they are in fact part of a dialogue does not represent a real-world end-use scenario.
An alternative evaluation setup is proposed in which raters interact with an avatar.
The experiments on conversational data in Section~\ref{sec:results_conversational} follow this work, in the sense that turns in the dialogue are presented in context rather than in isolation.
A key difference is that we do not incorporate an interactive setting. This allows for comparison between the three different settings we propose, none of which involve interaction.

In \cite{hu2016discourse} discourse structure is taken into account for improving prosody of longer passages of text.
The focus in this work, however, is on the improvements of a supervised signal pertaining to rhetorical structure, rather than on the evaluation. 

It is observed in \cite{latorre2014speech} that evaluating sentences in isolation ``may not be appropriate to measure the performance of intonation models.''
However, the objective in \cite{latorre2014speech} is to show that when evaluating single sentences without providing context, multiple prosodic variants of the same sentence might be equally valid according to raters.
No experiments were done to determine how those ratings change if a context is provided.  

Lastly, an evaluation protocol for an audiobook reading task, adapted from the scales proposed by \cite{p85}, is presented in \cite{hinterleitner2011evaluation}.
The method is aimed at a fine-grained analysis of the audiobooks task in particular, and does not cover an analysis of different evaluation alternatives.

In short, to the best of our knowledge, no systematic analysis of the effect of different ways to evaluate long-form \tts context has been carried out before.
The absence of such investigation is the primary motivation for this study.

\section{Evaluating Long-form \tts}
\label{sec:ways}

\begin{figure*}[ht]
  \centering
  \includegraphics[width=0.75\textwidth]{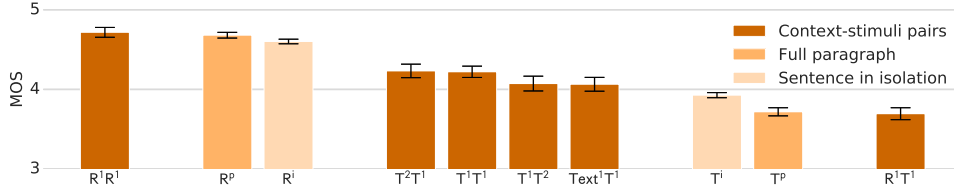}
  \caption{MOS results on the news reading data set across evaluation strategies. 'R' refers to real speech, 'T' is for TTS (synthesized speech), 'Text' means no speech but text. For evaluations without context, superscript 'p' denotes a full paragraph and superscript 'i' denotes sentences in isolation. For evaluations with context, '\rOne\rOne' is a context-stimulus pair of one line of real speech context and one line of real speech stimulus, '\tTwo\tOne' is two lines of TTS context, one line of TTS stimulus.}
  \label{fig:mos_all}
\end{figure*}

We present three ways to evaluate long-form material: as single sentences in isolation, as full paragraphs, and as context-stimuli pairs.
We should note that, even as the discussion below is presented in terms of sentences in a paragraph, it equally applies to turns in a dialogue.
Furthermore, although the discussion is applied to \mos, it is independent of what type of evaluation is performed and applies equally to \sxs tests as well as other varieties of evaluation such as MUSHRA \cite{mushra}.

\subsection{Evaluating sentences in isolation}

Firstly, we can use the traditional \tts approach and evaluate individual sentences separately as if they were isolated sentences. As mentioned in Section~\ref{sec:introduction}, the obvious disadvantage of this approach is that in evaluating isolated sentences, we are not considering the fact that these sentences are part of a larger discourse which may affect the way they should be synthesized.
There are, however, advantages to this method of presentation.
Raters, for example, are less likely to be able to infer the content based on context, in this setting, so lack of intelligibility is more likely to result in bad naturalness scores. 

In the work presented here we treat this method of evaluation as a reference to compare other results to, which allows us to determine empirically whether we learn something different using alternative evaluation methods. 

\subsection{Evaluating full paragraphs}

At the other end of the scale is the evaluation of full paragraphs. Evaluating full paragraphs imposes a higher cognitive load on raters which may impact the responses obtained.
Paragraph length, becomes an issue in its own right, and we may get different results depending on how long the paragraphs are.
An advantage of this setting, however, is that it is possible for raters to make judgments on the overall flow of the sentences in the paragraph, something they cannot do when they hear them in isolation.

\subsection{Evaluating context-stimulus pairs}

To compromise between evaluating isolated sentences and paragraphs we can present one or more sentences of the paragraph as context to the rater, and the subsequent sentence or sentences as the stimulus to be rated.

This approach raises questions regarding the amount of material that should be presented, both as context and as stimulus. 
Should we constrain the length of the context and stimulus in terms of the number of sentences or by overall length in words or syllables? E.g, 
a single long sentence may be longer than two short sentences.
In the work presented here, we choose to control the variation in terms of number of sentences and length of paragraphs.
We also evaluate whether paragraph length influences paragraph MOS scores (see Section~\ref{sec:paragraph_vs_sentence_ratings}).

Figure~\ref{fig:eval_isolated_with_context} shows various options for contexts.
To keep the figure clear a single sentence stimulus is shown, but we note that multiple sentences can be presented as a stimulus too.

\section{Experimental Setup}
\label{sec:experimental_setup}

We compare the three approaches for long-form TTS evaluation outlined above: 1) sentences in isolation, 2) full paragraphs 3) context-stimulus pairs. 

To test for consistency across different domains, we present results of evaluations in two distinctly different scenarios: news-reading and read conversations.
We use a WaveNet \cite{oord2018parallel} TTS voice that was built incorporating in-domain training data.

\subsection{Data and TTS system}
\label{sec:tts_system}
For our first series of evaluations we use a proprietary data set of read news articles.
We select only paragraphs containing two sentences or more.
Single sentence paragraphs are less interesting for our evaluations as any evaluation comparing single sentences to one-sentence paragraphs will come out even.
In our final dataset, we have 103 paragraphs of news material.
The longest paragraph contains 9 sentences, and the mean length is 3.0 sentences. To further compare the contexts of different lengths, we select a subset of paragraphs with a minimum length of three sentences resulting in a subset of 57 paragraphs with a mean length of 3.8 sentences.

The second data set consists of read conversations where two speakers take turns speaking.
We use turns in conversation as the units making up our stimuli (similar to using sentences in paragraphs in the previous setting).
An individual turn itself may consist of multiple sentences, which we keep together as a single turn.
The conversation design determines the total amount of variation of length per turn and keeps the amount of material per turn reasonably balanced.
We use two pairs of speakers.
The first pair recorded 42 conversations and the second pair recorded 71.
A key difference between this dataset and the news reading one is that the speaker changes between turns.

We should note that for the first dataset, a held-out set of passages was used for evaluation.
In the conversation case we did not have sufficient data to do this, and the conversations used for evaluation were used as training example for the WaveNet voice as well.
This is suboptimal, but as we are not trying to assess how well a particular \tts model can generalize, this should not affect the results presented here---we have seen little evidence that WaveNet models over-fit in such a way that any one utterance can have a significant impact on the resulting voice.

To synthesize speech we use a two-step approach where one model is trained to produce prosodic parameters ($F_0$, $c_0$ and duration) \cite{vwan2019chive} to be used by a version of WaveNet \cite{oord2018parallel}, trained separately to produce speech from linguistic features and the predicted prosodic parameters.
The model is not context-aware; it synthesizes speech sentence by sentence. 

\subsection{Rating task}
We use a crowd-sourced  MOS rating task for evaluation, where raters are asked to rate naturalness for the settings that do not include context, and appropriateness where the stimulus follows a context.
Stimuli are rated on a scale of 1-5.
The whole number points of the scale are labeled 'poor', 'bad', 'fair', 'good' and 'excellent'.
Raters are allowed to rate at 0.5 increments of the scale, as we find this gives slightly finer resolution in MOS scores at the top end of the scale. Stimuli are presented to raters in blocks of 10, except for the full paragraphs, which are presented in blocks of 5. Each stimulus is presented 8 times per experiment to randomly assigned raters and the MOS results presented are calculated from the averages of those 8 ratings for each stimulus.
Raters not using headphones are omitted from the analysis. The number of raters per task varies due to the overall number of stimuli in the task, with the lowest number of raters in a task being 35.

\subsection{Evaluation tasks}

\begin{figure*}[t]
  \centering
  \includegraphics[width=0.978\textwidth]{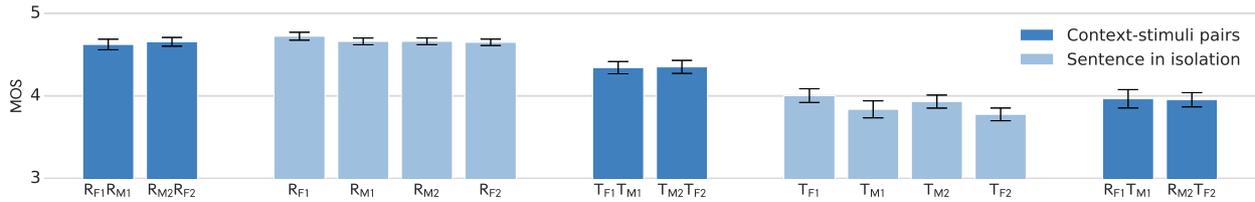}
  \caption{MOS results on the conversational data set, presented with context (F1 M1 and M2 F2) and in isolation (F1, M1, M2, F2). }
  \label{fig:mos_conv}
\end{figure*}

For \textbf{news reading} the following evaluations are carried out:

\begin{enumerate}[leftmargin=3ex]
    \item \textbf{Sentences in isolation} Both real speech and \tts versions of each sentence are presented as stimulus. Below, these results are referred to as \ri\xspace (Real speech, individual sentences) and \ti\xspace (\tts, individual sentences).
    
    \item \textbf{Full paragraphs} The same data is used as above, but presented as full paragraphs. Both real speech and \tts versions are presented to the raters. These results are labelled \rp\xspace and \tp, respectively.
    
    \item \textbf{Context-stimulus pairs} The first and second lines of paragraphs are presented, where the first line is the context and the second line is the stimulus to be rated. We experiment using a combination of real speech, \tts and text as the context, and both real speech and \tts as the stimulus.    Additionally, to evaluate varying the length of the context, we provide two lines either as context or as stimulus. In this setting, only \tts is used as context:
    
    \begin{description}
        \item[\rOne\rOne] One sentence real speech as context, one sentence real speech as stimulus;
        \item[\rOne\tOne] One sentence real speech context, one sentence \tts as stimulus;
        \item[\tOne\tOne] One sentence \tts context, one sentence \tts stimulus; 
        \item[\textOne\tOne] One sentence textual context, one sentence \tts stimulus.
        \item[\tTwo\tOne] Two sentence \tts context, one sentence \tts stimulus;
        \item[\tOne\tTwo] One sentence \tts context, two sentence \tts stimulus;
    \end{description}
\end{enumerate}
In the news reading tasks, real speech samples \ri\xspace are cleaned from sentence-initial breathing noise and \rp\xspace from paragraph-initial breathing noise. All real speech samples are downsampled to match the TTS sampling rate.

The \textbf{conversational} data includes two pairs of speakers: \textbf{F1} paired with \textbf{M1}, and \textbf{M2} paired with \textbf{F2}, where \textbf{F} and \textbf{M} denote female and male speakers, respectively. The evaluations use \tts samples and real speech samples from all four speakers: \tfOne, \tfTwo, \tmOne\xspace, \tmTwo\xspace and \rfOne, \rfTwo, \rmOne\xspace, \rmTwo, respectively.

WaveNet voices were built for each of these speakers.

\section{Results}
\label{sec:results}

In this section we discuss the results of the two sets of experiments performed.
We use a two-tailed independent t-test with $\alpha=0.05$ for calculating significance between results.

\subsection{News reading}

\begin{table*}[t]
    \small
    \centering
    \caption{Correlations of sentence MOS scores with paragraph MOS (news reading data).}
    \begin{tabular}{c|p{1.3cm}|p{1.3cm}|p{1.3cm}|p{1.3cm}|p{1.3cm}|p{1.3cm}||p{1.3cm}|p{1.3cm}}
      \textbf{Correlate} & \textbf{Mean sentence MOS}   &  \textbf{First Sentence MOS}  &  \textbf{Second Sentence MOS} & \textbf{Last sentence MOS}  &
      \textbf{Min. sentence MOS}  & \textbf{Max. sentence MOS} & \textbf{Paragraph no. of sentences} & \textbf{Paragraph no. of words}\\
      \hline
         \textbf{r} &
         \textbf{0.296} &
         0.087 &
         0.114 &
         \textbf{0.268} &
         \textbf{0.234} &
         \textbf{0.345} &
         -0.020  &
         0.029 \\
         \textbf{p} &
         \textbf{$<$ 0.05} &
         $>$ 0.05 &
         $>$ 0.05 &
         \textbf{$<$ 0.05} &
         \textbf{$<$ 0.05} &
         \textbf{$<$ 0.01} &
         \textbf{$<$ 0.05} &
         $>$ 0.05 \\
    \end{tabular}
    \label{tab:correlations}
\end{table*}

\begin{table}[t]
    \small
    \centering
    \caption{Regression model coefficients for predicting the paragraph MOS from individual sentence MOS for paragraphs of lengths two, three and four sentences long (news reading data).}
    \begin{tabular}{ccccc}
    \multicolumn{5}{c}{\textbf{Model for paragraphs of two sentences}}\\
    \multicolumn{3}{c}{Num. of paragraphs} & \multicolumn{2}{c}{46} \\
    \multicolumn{5}{c}{$R^2 = 0.04, \quad(F = 0.95,\ p > 0.05)$} \\
\hline
        & coef   &  std err     & t   & $P{>}|t|$   \\
intercept   &   2.50  &  0.92  &    2.74  &    0.01 \\
s1      &       0.15  &  0.15  &    0.99  &    0.33 \\
s2      &       0.17  &  0.15  &    1.13  &    0.26 \\
\hline\hline
    \multicolumn{5}{c}{\textbf{Model for paragraphs of three sentences}}\\
    \multicolumn{3}{c}{Num. of paragraphs} & \multicolumn{2}{c}{31} \\
    \multicolumn{5}{c}{$R^2 = 0.27, \quad(F=3.35,\ \mathbf{p < 0.05})$} \\
\hline
        & coef   &  std err     & t   & $P{>}|t|$   \\
intercept   &   1.30  &    0.95  &    1.37  &    0.18   \\
s1          &   0.39  &    0.14  &    2.77  &    0.01   \\   
s2          &   0.01  &    0.12  &    0.11  &    0.92   \\ 
s3          &   0.22  &    0.14  &    1.63  &    0.17   \\
\hline\hline
    \multicolumn{5}{c}{\textbf{Model for paragraphs of four sentences}}\\
    \multicolumn{3}{c}{Num. of paragraphs} & \multicolumn{2}{c}{15} \\
    \multicolumn{5}{c}{$R^2 = 0.54, \quad(F = 2.97,\ p > 0.05)$} \\
\hline
        & coef   &  std err     & t   & $P{>}|t|$   \\
intercept   &   4.23  &    2.03   &   2.08  &    0.06   \\   
s1      &      -0.53  &    0.25   &  -2.14  &    0.06   \\   
s2      &       0.12  &    0.16   &   0.73  &    0.48   \\   
s3      &       0.17  &    0.17   &   1.00  &    0.34   \\  
s4      &       0.11  &    0.22   &   0.51  &    0.62   \\   
    \end{tabular}
    \label{tab:regression}
\end{table}

Figure~\ref{fig:mos_all} shows the results for all MOS evaluations, ordered from high to low.

The first block of results confirms the intuition that real speech scores higher than all settings involving \tts.
The highest ratings are for appropriateness of a real speech stimulus in a real speech context (\rOne\rOne).
The scores are slightly higher than for naturalness ratings of both real speech paragraphs (\rp) and real speech isolated sentences (\ri).
Real speech paragraphs (\rp) themselves are rated slightly higher than real speech isolated sentences (\ri).
Within this grouping of real speech results, there are significant differences between all three conditions.
These results alone already indicate that there is a difference between evaluating sentences in isolation and in context, even when only real speech is involved. 

The next block of results in Figure~\ref{fig:mos_all} shows the results for context-stimuli pair evaluations.
Presenting two sentences as context, while rating one follow-on sentence (\tTwo\tOne) scores highest, followed by one sentence as context with one sentence rated (\tOne\tOne).
The lowest scores are obtained when one sentence is presented as context followed by two sentences being rated (\tOne\tTwo). \tOne\tTwo\xspace is found to be significantly different from \tOne\tOne and \tTwo\tOne.
The final bar in this block shows the result of presenting the context as text rather than speech (\textOne\tOne) and this gives a  score not significantly different from \tOne\tTwo.
These results indicate that the length of the context presented does not appear to have a significant effect on the MOS results, but increasing the length of the stimulus lowers the MOS result.

The next block holds the results for evaluating \tts sentence (\ti) and paragraph (\tp) naturalness in isolation. These results are significantly lower than the ones in the previous blocks.
One potential explanation why raters would rate paragraphs lower than their individual sentences, is that ratings are strongly influenced by the worst thing they hear in the stimulus and thus as the stimulus becomes longer the rating is likely to be lower. This interpretation is consistent with the result above where a lower MOS was found when increasing the stimulus length in a context.
It could suggest a (weak) correlation between the minimum sentence MOS and the paragraph MOS (cf. Table~\ref{tab:correlations} discussed in Section~\ref{sec:paragraph_vs_sentence_ratings}).
Alternatively, it may be that higher cognitive load simply results in lower ratings.

It is interesting to see that sentences with context are rated higher than when presented in isolation.
As noted in Section~\ref{sec:tts_system}, the \tts model used is not taking any paragraph level context into consideration, so the difference has to be attributed either to the task itself, or the fact that the content of a paragraph non-initial sentence sounds less natural when presented out of context.

The final and lowest result in Figure~\ref{fig:mos_all} is for \tts stimuli with real speech context (\textbf{\rOne\tOne}).
A key observation to point out is that these results are considerably lower than the ones where \textit{the same stimuli} are presented with a \tts context.
This seems to indicate an anchoring effect of the real speech lowering the perceived quality of the TTS, suggesting that when rating appropriateness in context, raters pay particular attention to whether the quality of the stimulus matches the quality of the context.\newline
Lastly, the fact that cases where the \tts context was used score higher than when sentences are rated in isolation suggests that part of the appropriateness judgment relates to similarity in quality compared to the context, and the rating does not just relate to overall naturalness and how well the prosody is suited in context to the paragraph.
The implication here is that the context-stimuli setting cannot be considered to be an alternative to the sentence-in-isolation naturalness MOS task, because it will produce varying results depending on the quality of the context.
The MOS result a context-stimulus evaluation yields can be substantially higher than one obtained for a sentence in isolation when there is a quality match between the context and stimulus, or lower it when the quality of the context is higher than that of the stimulus.

\subsection{Conversations}\label{sec:results_conversational}

To determine if the differences observed between ratings for sentences presented in isolation versus sentences presented in context are consistent across domains, we perform evaluations on a distinctly different dataset that consists of conversations.
We restrict the evaluation to using only the first and second turns of the dialogues, as we saw previously that amount of context presented did not greatly affect the results.

We both evaluate the first and second turns in isolation, and we evaluate the second turns using the first turns as context.
Note that, different from the news data, the context in this scenario is uttered by a different speaker. 
In two separate tasks, we present the context either as real speech or as \tts. 

The results of this experiment are shown in Figure~\ref{fig:mos_conv}.
The MOS for the evaluations involving only recorded voices \rfOne, \rfTwo, \rmOne\xspace and \rmTwo\xspace range between 4.6~--~4.7 with no statistically significant difference between the scores for the second turns presented in isolation or in their recorded context---the only statistically significant difference in this group of evaluations is observed between \rfOne\xspace and \rfOne\rmTwo\xspace or \rfTwo.
Furthermore, MOS scores for the synthesized turns in isolation range from 3.8 for voices \tmOne, \tmTwo\xspace and \tfTwo\xspace and to 4.0 for voice \tfOne.
Conversely, when the second turns of the dialogues are preceded by their context, the MOS for the \tts voices rises to the 4.3~--~4.4 range, mirroring the effect we saw for the read news data.
Furthermore, using real speech as context (\rfOne\tmOne\xspace and \rmTwo\tfTwo) decreases the resulting MOS for \tts stimuli as again the raters appear to consider the quality of the context as an anchor. However in this case these ratings do not drop below the ratings of the turns in isolation.
We attribute this to the fact that, even if the context is presented as the natural speech of a different speaker, this still acts as an anchor, but a weaker one than the natural speech of the same speaker would be.

\section{Further analysis}\label{sec:further_analyis}

The results presented in the previous section show that rating a full paragraph gives different results then rating sentences in isolation does, regardless of how the task is set up.
To gain more insight into this observation, we analyze correlations between full paragraph and sentence ratings. These tests are carried out on the news reading data set.

\subsection{Correlating ratings of full paragraphs and single sentences}
\label{sec:paragraph_vs_sentence_ratings}

Table~\ref{tab:correlations} shows the correlations (Pearson's $r$) between paragraph MOS scores and various sentence MOS ratings.
We see significant correlations, at the 5\% level, of around 0.3 between the MOS rating of the full paragraph and the 'Mean sentence MOS', 'The last sentence MOS' and the 'Minimum sentence MOS'.
Furthermore, there is a significant correlation of 0.345, at 1\% level, between paragraph MOS and the maximum MOS of the sentences the paragraphs consists of.
All of these $r$ values are small, however, and only 12\% of the variance can be accounted by the maximum sentence MOS of $r$ = 0.345.
These correlations show that paragraph MOS is influenced by the individual sentences, both collectively through the means and individually through the extremes, yet only less than half the variance can be accounted for this way.
We conclude that, while paragraph and individual sentence ratings cannot be considered to be independent, the majority of the variance seen in paragraph MOS scores is not accounted for by the MOS ratings of the individual sentences.

Lastly, the rightmost columns of Table~\ref{tab:correlations} shows the correlations between the paragraph length (measured in sentences or words) and the paragraph MOS rating. There is a correlation, supporting the intuition that MOS ratings go down as paragraphs get longer in terms of sentences, but the $r$ value is so small that it does not appear to be meaningful.

\subsection{Correlating ratings of full paragraphs and single sentences and their positions}

An alternative hypothesis is that, even if little correlation between the MOS ratings of paragraphs and the MOS ratings of each individual constituent sentence is found when these are analyzed all together, perhaps the latter can be inferred from the former if the order of sentences is taken into account.
To test this hypothesis we create linear regression models predicting the paragraph MOS from the individual sentence MOS values, depending on their position in the paragraph.
We restrict these experiments to paragraphs of length two, three and four sentences, as we do not have sufficient data in the current experiments to analyze longer paragraphs, and it is not immediately clear how models allowing for variable paragraph length should be designed.

The results of the regression experiments are shown in Table~\ref{tab:regression}. First, we note that the only model with a significant $R$-squared value, i.e., that can account for a variance in a significant way, is the model for paragraphs of three sentences.
For this model the only significantly non-zero contribution is made by the MOS of the first sentence.
That trend is not repeated for the other models.

For the two sentence model only the constant term contributes in a non-zero way, i.e., the paragraph MOS is the same for all paragraphs under this model.
Hence, it is unsurprising that this has an insignificant $R$-squared value. 

For paragraphs of four sentences there are non-zero contributions from both the constant term and the first sentence in the paragraph, but the low and non-significant $R$-squared value for this model means this model does not fit the data well.

In short, we conclude from these results that the individual MOS ratings of sentences are bad predictors of paragraph MOS, and if a MOS which reflects the overall quality of the paragraph is required, it needs to be obtained directly.

\section{Conclusions}
\label{sec:conclusions}

Now that the performance of \tts systems has come to a level where voice quality itself is close to human level, interesting and challenging new tasks are being undertaken, like synthesizing speech for an entire audio book or in a multi-turn conversation.
The experiments presented here suggest that, as these new tasks go beyond the scope of traditional \tts, new ways of evaluation should be considered including task based evaluations.

We demonstrated that long-form evaluation can be improved beyond evaluating isolated sentences by showing that different results are obtained when the material is presented in different ways.
Asking raters to rate the paragraph as a whole does not give the same results as asking raters to rate the constituent sentences in isolation or asking raters to rate using the previous parts of the paragraph as context. Additionally, we proved that it is difficult and inconclusive to try to predict paragraph MOS from the MOS of the individual sentences in it, which suggests that raters do pay attention to contextual cues when performing these different tasks.

We conclude, therefore, that to fully evaluate long-form paragraphs or dialogues, a combination of tests is necessary.
In some circumstances it may be sufficient to only evaluate the paragraphs as a whole, and this is probably what should be done if resources are limited and the paragraphs are not too long.
Yet, as observed above, this method gives lower scores than the scores for individual sentences when rated either in isolation or with discourse context.
One potential reason is that although our \tts training data consists of multi-sentence data, no significant effort has been made to model paragraph level structures in a \tts system, for example varying the prosody of a sentence based on the content or realization of the previous sentence, and it will be interesting to see if successfully doing so can close the gap between the rating for paragraphs and sentences.

One shortcoming of the three different approaches of evaluating long-form \tts we presented is that they do not consider unbalanced numbers of sentences per paragraph in the data.
That is, we have a lot more second sentences in a paragraph than we do fifth sentences in a paragraph.
Future work could investigate how to handle unbalanced data in a rigorous way.

Lastly, evaluating sentences in context produced interesting results with higher scores in general, specifically when the context was also \tts: with the same voice in the case of the read news experiments, but also when the context was a different \tts speaker, in case of the conversation experiments.
We attribute this effect to the raters including a similarity judgment between the quality of the context and stimulus in their scores.
This is corroborated by the experiments with real speech context, which yielded lower ratings.
Evaluating in context is therefore our recommended way to evaluate long-form material as it allows sentences to be presented individually, while paragraph effect judgments can be considered in the rating.

\section{Acknowledgments}
We would like to acknowledge contributions to this work from the wider \tts research community within Google AI and DeepMind, your thirst for understanding lead to this study. Specific thanks to Xingyang Cai, Anna Greenwood, Mateusz Westa, Dina Kelesi and Leilani Kurtak-McDonald for help with evaluation tools and voice building.  

\bibliographystyle{IEEEtran}
\bibliography{main}

\end{document}